\documentstyle[preprint,eqsecnum,aps]{revtex}
\tighten
\newcommand{\beq}{\begin{equation}}
\newcommand{\beqa}{\begin{eqnarray}}
\newcommand{\eeq}{\end{equation}}
\newcommand{\eeqa}{\end{eqnarray}}
\newcommand{\simg}{\gtrsim}
\newcommand{\siml}{\lesssim}

\begin{document}
\draft

\preprint{YITP-98-7, WU-AP/70/98, KUNS-1490}

\title{Two Boosted Black Holes in  Asymptotically de Sitter
Space-Time\\[.3em]
 - Relation between Mass and Apparent Horizon Formation -}

\author{Takeshi Chiba$^{1}$, Kei-ichi Maeda$^{2}$,
Ken-ichi Nakao$^{3}$, Taro Tsukamoto$^{4}$}

\address{$^{1}$Yukawa Institute for Theoretical Physics,
Kyoto University,  Kyoto 606-01, Japan.}
\address{$^{2}$Department of Physics, Waseda University,
Tokyo 169, Japan.}
\address{$^{3}$Department of Physics, Kyoto University,  Kyoto
606-01,  Japan.}
\address{$^{4}$National Aerospace Laboratory, Tokyo 182,
Japan.}

\date{\today}

\maketitle

\begin{abstract}
 We study the apparent horizon for two boosted
black holes  in the asymptotically de Sitter space-time by
solving the  initial data on a space with punctures.
We show that the apparent horizon enclosing both black holes
is not formed if the conserved mass of the system
(Abbott-Deser mass) is larger than a critical mass.
The black hole with too large
AD mass therefore cannot be formed in the asymptotically  de
Sitter space-time even though each black hole has any inward
momentum.  We also discuss the dynamical meaning of AD mass
by examining the electric part of the Weyl  tensor (the tidal
force)  for various initial  data.

\end{abstract}

\section{Introduction}

The inflationary universe scenario is one of the most
favorable model  to explain the present isotropy and
homogeneity of the  Universe\cite{inf}.  The basic idea of this
scenario is that  the potential energy of a scalar field, which
behaves as an effective cosmological constant, dominates and
causes the de Sitter-like rapid cosmic expansion in the early
stage of the Universe.
Then it seems to be likely that due to this rapid cosmic
expansion  the initial anisotropy and inhomogeneities might be
stretched out and  the universe becomes homogeneous and
isotropic.  Such a picture is based on the so-called cosmic no
hair conjecture (CNHC)  which states that ``all" space-times
with a cosmological constant $\Lambda$  approach  the de Sitter
space-time asymptotically\cite{gibbons}.  Of course, CNHC is
not always true without any additional conditions since we know
that  some inhomogeneities can gravitationally collapse into
black holes  in the space-time with $\Lambda$.  Hence the
dynamics of the inhomogeneities is an important  issue to get
physical insight into the present homogeneity and  isotropy of
the Universe and there are many researches  on this
problem\cite{gp}.

Recently, we have studied
numerically apparent horizons in the initial data in the
asymptotically de Sitter space-time \cite{nmno} - \cite{cm}.
The results suggest that there is an upper limit on
the gravitational mass, the Abbott-Deser (hereafter AD)
mass \cite{ad} of a black hole.
The AD mass is the corresponding notion to the ADM mass
in the asymptotically flat space-time.
Hence large inhomogeneities may not collapse into a black
hole and,  furthermore, large black holes may not collide
in the asymptotically de Sitter space-time.
The same result was obtained by the analysis of the
Oppenheimer-Snyder model with $\Lambda$\cite{dust}.
Further, dynamical  simulations
for  the Brill waves in the asymptotically de Sitter space-time
were performed and  revealed the same results as the spherically symmetric
dust  collapse\cite{SHIBATA}.
It was also shown that under some conditions there is
an upper limit on the area of the event horizon in the
asymptotically de Sitter space-time\cite{hsn}.
Since, as in the case of the asymptotically flat  space-time,
the area of the black hole does not decrease in the
asymptotically  de Sitter space-time, the black holes with too
large total area of the  event horizons may not collide and
not  merge if the cosmic censorship hypothesis holds.
 We note,
however, that the relation between the  AD mass and the event
horizon formation is not yet clear in the  asymptotically de
Sitter space-time.

Although the above analysis show some clear relation between the
AD mass and apparent horizon formation, there   still
remain some unanalyzed effects, one of which is initial motion
of a black hole in two-black hole system. If two black holes
have inward velocity, we may expect that those   black holes
might collide each other and will form a single black hole.
Then,  in this paper, we shall make a further study of the AD
mass in  two-black-hole system.
Previous work about two-Einstein-Rosen bridge system was
focused only on the  cases with no relative velocity other than
the uniform background cosmic  expansion\cite{nym}.
Hence, in order to see the effect of the relative velocity,
we shall investigate the axisymmetric initial data
of two non-spinning black holes with finite velocity in
addition to
 the background cosmic expansion. We solve the initial  data on
a space  with punctures following \cite{bb}.
Then we search for the apparent horizon enclosing both  black
holes  in order to get some insight into the dynamics of the
inhomogeneities  in the inflationary universe.
We find that such an apparent horizon does not appear  no
matter how close together each black hole is and how fast it
moves   when AD mass  of this system is larger than a critical
value which agrees  with that of the  Schwarzschild-de Sitter
space-time\cite{carter}.

It is worth noting  that as pointed out in \cite{NSM}, 
AD mass can vanish and furthermore
can be negative even if inhomogeneities
satisfy energy conditions.
In this article, we shall explicitly show this fact in the case of the
initial data containing two black holes.
So a question arises:
``Is gravitational field produced when AD mass vanish?"
Then, in order to see the dynamical meaning of AD mass,  we
 investigate the asymptotic behavior of the electric part
of  the Weyl tensor which corresponds to the tidal force.
In contrast with the asymptotically flat space-time, the tidal
force comes from  the anisotropic velocity  as well as AD mass.
In particular, the contribution from the anisotropic velocity
becomes the  leading term in the asymptotic region when the
conserved momentum does not vanish. When AD mass
vanishes, the tidal force is   produced solely by the
anisotropic momentum.  However, even if the dynamical effects
of the anisotropic velocity on  the large scale inhomogeneities
might be important,  the apparent horizon formation is
essentially determined  by the AD mass and furthermore the
critical mass coincides with  that of the Schwarzschild-de
Sitter space-time.  This fact suggests that the final state of
non-rotating gravitational  collapse in the asymptotically de
Sitter space-time might be the  Schwarzschild-de Sitter
space-time if naked singularities are not  formed. Hence, it
seems that the inhomogeneities with too large AD mass  cannot
collapse into a single black hole.

This paper is organized as follows.
In Sec.II, we present initial data of two masses with
linear momenta in the asymptotically de Sitter space-time and
search for the apparent horizon, numerically.
In Sec.III, we investigate the electric part of the Weyl
tensor of the  conformally flat initial data and discuss the
effect of AD mass  and
the linear momentum on the electric part of the Weyl tensor.
Then Sec.IV is devoted to discussion.
We adopt the units of $c=G=1$. Our notations follow
those of MTW\cite{mtw}.

\section{Initial Data for Two Boosted Masses
in the Asymptotically de Sitter Spacetime}

\subsection{Setting up Initial Data}

In order to obtain initial data, we have to solve
the Hamiltonian and momentum constraints,
%
\beqa
   {^{(3)}}R-{\hat K}_{i}^{j}{\hat K}_{j}^{i}+{2\over3}K^{2}
   &=&6H^{2}+16\pi \rho_H,
\label{const:ham}\\
   D_{j}({\hat K}_{i}^{j}-{2\over3}\delta_{i}^{j}K)& = &8\pi
J_i,
\label{const:mom}
\eeqa
%
with
%
\beq
 H \equiv \sqrt{\Lambda \over 3},
\eeq
where ${^{(3)}}R$ is the Ricci scalar of the 3-space.  ${\hat
K}_{i}^{j}$  and $K$ are the traceless part and the trace of
the  extrinsic curvature of the 3-space, respectively. $D_{j}$
is  the covariant derivative with respect to the intrinsic
metric of 3-space.
$\rho_{H}$ and $J_{i}$ are, respectively, the energy density
and  the momentum density of the matter fluid measured by the
normal line  observer.

In order to solve Eq.(2.1), we follow the York's conformal
prescription\cite{york} and
assume the conformally flat metric,
%
\beq
d\ell^{2}=\psi^{4}(R,z)f_{ij}dx^{i}dx^{j},
\label{cflat}
\eeq
%
where $f_{ij}$ is the flat Euclidean metric. We
set the trace part of the extrinsic curvature $K$ to
%
\beq
 K=-3H.
\label{cmc}
\eeq
%
Note that the condition Eq.(\ref{cmc}) turns out to be the
Friedmann equation $({\dot a}/a)^{2}=H^{2}$ for the scale factor $a$ in
the homogeneous and isotropic universe.
Hence, the condition (2.5) is regarded as the assumption of the
uniformly expanding background universe.

Then Eqs.(2.1) and (2.2) become as
%
\beqa
  {\tilde \Delta}\psi
   &=&-{1\over8}{\tilde K}_{i}^{j}{\tilde K}_{j}^{i}
\psi^{-7}-2\pi
\tilde{\rho_H}\psi^{-3},
\label{laplace}\\
  {\tilde D}_{j}{\tilde K}_{i}^{j}&=& 8\pi \tilde{J}_{i},
\label{conf:mom}
\eeqa
%
where ${\tilde K}_{ij}={\hat K}_{ij}\psi^{2},
\tilde{\rho_H}=\rho_H\psi^8, \tilde{J}_i=J_i\psi^6$, and
${\tilde D}_{i}$ is the covariant derivative with respect
to the flat metric $f_{ij}$ and ${\tilde \Delta}$ is the
Laplacian  operator of the flat space.
The traceless part of the extrinsic curvature  ${\tilde
K}_{ij}$ is  decomposed into the transverse traceless part and
the longitudinal  traceless part
%
\beq
 {\tilde K}_{ij}={\tilde K}_{ij}^{TT} +({\tilde L}{\tilde
W})_{ij},
\label{kij}
\eeq
%
with
%
\beq
({\tilde L}{\tilde W})_{ij}\equiv  {\tilde D}_{i}{\tilde W}_{j}
    +{\tilde D}_{j}{\tilde W}_{i}
    -{2\over3}f_{ij}{\tilde D}_{\ell}{\tilde W}^{\ell}.
\label{def:lw}
\eeq
%
As for the transverse traceless part of the extrinsic curvature,
we  assume ${\tilde K}_{ij}^{TT}=0$. Restricting to
the vacuum case, i.e. $\tilde{\rho_H}=\tilde{J_i}=0$,  Eq.(2.7)
becomes
%
\beq
 {\tilde \Delta} {\tilde W}_{i} +{1\over3}{\tilde D}_{i}{\tilde
D}^{j}
    {\tilde W}_{j}=0.
\label{lw}
\eeq
%

The analytic  monopole solution for ${\tilde W}^{i}$ is  given
by \cite{bowen}
\beq
{\tilde W}^{i}=-{1\over 3r}(7P^i+n^iP^kn_k),
\label{wi}
\eeq
where $n^i=x^i/r$. $P^i$ is the linear momentum defined by
\beq
P^i={1\over 8\pi} \oint \tilde{K}^{ij}d^2\tilde{S}_j,
\eeq
where $d^2{\tilde S}_{i}$ is the area element of the sphere in
the  conformal flat space.
Just for simplicity, we consider only the initial data of  two
black holes with equal mass  and equal magnitude of momentum.
Since Eq.(\ref{lw}) is linear, we can easily find
the extrinsic curvature for the two-black-hole system with
linear momenta by superposition as
\beq
{\tilde K}_{ij}=K_{ij}^{(+)}+K_{ij}^{(-)},
\label{Kij}
\eeq
where $K_{ij}^{(\pm )}$ comes from each
hole  located at $(0,0,\pm a)$.
Then $K^{(\pm ) ij}$   is  given  as
\beq
K^{(\pm ) ij}={3\over 2r_{\pm}^2}\left[P^{(\pm ) i} n^{(\pm )
j} +P^{(\pm ) j} n^{(\pm ) i} -(f^{ij}-n^{(\pm ) i}n^{(\pm
) j})P^{(\pm ) k}n^{(\pm )}_k\right],
\label{kij:sol}
\eeq
where $n^{(\pm ) i}$ and $P^{(\pm ) i}$ are defined by
\beqa
 n^{(\pm ) i}&=&{1 \over r_{\pm}}(x,y,z \mp a), \\
P^{(\pm ) i}&=&(0,0,\pm P),
\eeqa
with $r_{\pm}=\sqrt{x^2+y^2+(z\mp a)^2}$. Substituting ${\tilde
K}_{ij}$ obtained above  into Eq.(\ref{laplace}),
we get the equation for the conformal factor $\psi$.  We
rewrite the conformal factor following \cite{bb}
\beq
\psi = {M_0\over 2r_+}+{M_0\over 2r_-} +u\equiv {1\over
\alpha}+u,
\label{u}
\eeq
where $M_0$ is a constant.
Then the Hamiltonian constraint becomes
\beq
 {\tilde \Delta}u
   =-{1\over8}\alpha^7{\tilde K}_{i}^{j}{\tilde K}_{j}^{i}(1
+\alpha u)^{-7}.
\label{laplace2}
\eeq
It should be noted that the coefficient in Eq.(\ref{laplace2})
is no longer singular as opposed to Eq.(\ref{laplace}) because
the monopole terms are subtracted by Eq.(\ref{u}).
The conformal factor $\psi$ (or $u$) is obtained by solving
iteratively   Eq.(\ref{laplace2}) by a finite difference method
typically  with $200 \times 200$ grid numbers in cylindrical
$(R,z)$ coordinate.  The boundary conditions are
given by
%
\beqa
\psi \rightarrow 1&+&{M \over 2r}~~~~~~~~~~{\rm for}~~
     r \equiv \sqrt{x^{2}+y^{2}+z^{2}} \rightarrow +\infty,
\label{asympt}\\
{\partial u\over \partial R}\bigg{|}_{R=0}&=&{\partial u\over
\partial z}\bigg{|}_{z=0}=0. \label{zero}
\eeqa
%
Note that $2M_{0}$ agrees with $M$ in the limit
$P^{i} \rightarrow 0$.
By virtue of the time slicing condition Eq.(\ref{cmc}),
the boundary condition Eq.(\ref{asympt}) guarantees that the
spacetime  is asymptotically de Sitter one.
In practice, however,
 numerical solutions cover only a finite region. We therefore
use the Robin boundary condition at the outer boundary
\beq
{\partial u\over \partial r}={1-u\over r}.
\eeq

\subsection{Finding Apparent Horizons}

Now we search for apparent horizons in the initial data
prescribed above.  A marginally trapped surface  is a closed
2-surface $S$ where  the expansion $\Theta $ of
future-directed  outgoing null vectors $\ell^{\mu}$ normal to
it vanishes\cite{trap}.  The apparent horizon is defined as the
outer boundary of a connected component of the trapped region.
The apparent horizon is believed to
agree with the marginally outer trapped surface by the proof
in \cite{trap}.  However the proof is correct only if the
apparent  horizon is smooth and hence
exactly speaking, the apparent horizon might not agree with the
marginally outer trapped surface\cite{KODAMA}.
In the situation considered here, however, we find a smooth
apparent  horizon and hence we may regard the
outer marginally trapped surface  as the apparent horizon.

Let $s^{\mu}$ be the outward-pointing spacelike unit normal
to $S$ and $n^{\mu}$ be the unit normal to a time slice, then
$\ell^{\mu}$ can be written as $\ell^{\mu}=n^{\mu}+s^{\mu}$
and thus
\beq
\Theta=D_{i}s^i+K_{ij}s^is^j-K=0
\label{expansion}
\eeq
on $S$.
For the conformally flat space Eq.(\ref{cflat}),
Eq.(\ref{expansion})  is rewritten as
\beqa
\Theta=&-&{r\over \psi^2(r^2+r^2_{\theta})^{3/2}}
\Bigl[r_{\theta \theta}+\left({r_{\theta}^3\over r^2}
+r_{\theta}\right)
\left(4{\psi_{\theta}\over \psi}+\cot \theta\right)
-r_{\theta}^2\left({3\over r}+4{\psi_r\over
\psi}\right)-2r-4r^2{\psi_r\over \psi}\Bigr]\nonumber\\
&+&K_{ij}s^is^j-K=0,
\label{ah}
\eeqa
where suffices $r, \theta$ denote those partial derivatives.
To find an apparent horizon, we solve Eq.(\ref{ah})
as a two-point boundary value problem with boundary  conditions
$r_{\theta}=0$ at $\theta=0,\pi/2$ following Sasaki  et
al\cite{sasaki}.  This method was also used recently by
Cook et al.\cite{cook}.

Likewise the asymptotically flat space-time,
when an apparent horizon forms, there always exists an  event
horizon enclosing it in the asymptotically de Sitter space-time
if  there is no naked singularity and the null convergence
condition  is satisfied\cite{SNKM}. Hence, by investigating the
existence of the apparent horizon in various initial data,
we obtain some insight into the evolution of inhomogeneities.

As in the case of the asymptotically flat space-time,
when the separation $a$ between each black hole is short
enough, the apparent horizon encompassing both black holes
appears (see Fig.1).  The existence of such an apparent horizon
means the merging of two  black holes and we shall  search for
such an apparent horizon in an asymptotically de Sitter space.

In order to control the strength of the inhomogeneities,
we vary $a$, $P^{\pm}_{i}$ and $M_{0}$.
However, these parameters are not conserved quantities of
this system.  In the asymptotically de Sitter space-time, there
are ten conserved  quantities associated with ten Killing
vectors of the background  de Sitter space-time.
In the initial data considered here,
only the AD mass associated with the timelike Killing
vector of the background de Sitter space-time is the
non-trivial  conserved Killing quantity. (Conserved total linear
and angular momentums vanish here).  Hence, we utilize the AD
mass as a measure characterizing the strength of
inhomogeneities in addition to $a$ and $P$.

In general, in conformally flat initial data with the  time
slicing  condition Eq.(\ref{cmc}), the AD mass $M_{AD}$ is
expressed as \cite{ad,NSM}
%
\beq
M_{AD}=M+\Delta M,
\label{admass}
\eeq
%
with
\beqa
     M &\equiv& -{1 \over 2\pi}
        \oint d^2{\tilde S}_{i}f^{ij}\partial_{j}\psi,
\label{mass}\\
     \Delta M &\equiv& {H \over  8\pi}
        \oint d^2{\tilde S}_{i}{\tilde K}_{j}^{i} {\tilde
\xi}^{j}.
\label{deltam}
\eeqa
Here the surface integral is taken over an infinitely
large sphere and  ${\tilde \xi}^{i}=(x,y,z)$
is the conformal Killing vector
orthogonal to the sphere. Although $M$ agrees with the
definition of  the ADM mass and is non-negative, $\Delta M$ can
be negative. For example,  for the initial data that we
consider, we obtain from Eq.(\ref{kij:sol}) using the Gauss's
theorem in the region ${\bf R}^3-\{{\rm two~ points}\}$
\beq
\Delta M= {H\over 8\pi}\int_{{\bf R}^3-\{{\rm
two~ points}\}} d\tilde{v}
\tilde{D}_i({\tilde K}_{j}^{i}{\tilde \xi}^{j})
+ 2HaP= 2HaP,
\eeq
where $d\tilde{v}$ is the volume element in the conformal
space and  we have made use of Eq.(\ref{conf:mom}) and the
conformal Killing equation
$({\tilde L}{\tilde \xi})_{ij}=0$.
Therefore, $M_{AD}$ can be negative for $P<0$ (inward
velocity) if
\beq
M < 2Ha|P|.
\label{crit:negative}
\eeq

We shall numerically investigate the critical separation
$a_{c}$  such that the apparent horizon surrounding black holes
disappears for
$a>a_{c}$. In Fig.2, we display $a_{c}/H^{-1}$ with respect
to $M_{AD}$  normalized by $M_{c}$ which is defined by
%
\beq
 M_{c} \equiv {1 \over \sqrt{27}H}.
\eeq
%
In fact, $M_{c}$ is the critical mass of the
Schwarzschild-de Sitter space-time.
The Schwarzschild-de Sitter space-time is characterized
by $M_{AD}$, and when $M_{AD}<M_{c}$, it contains a  black hole.
However, in the case of $M_{AD} \geq M_{c}$, the
Schwarzschild-de Sitter space-time contains no black hole.  As
a test of our numerical code, we searched for an apparent
horizon with
$a=0$ and $P=0$, which corresponds to Schwarzschild-de Sitter
space with $M_{AD}= 2 M_0$. For
$M_{AD}/M_c > 1.007$, an apparent horizon is not found.
Therefore, our numerical code has the accuracy of  less than
$0.7\%$.

In Fig.2(a), we set that $|P|/M_{0}$ is equal to 1. The white
square shows that  of $P>0$ while the black square corresponds
to that of
$P<0$.
Fig.2(b) is the same but for $|P|/M_{0}=5$.
In the asymptotically flat case, the critical separation
$a_{c}$ with fixed $P/M_{AD}$ is proportional to $M_{AD}$ (OK?)
and  increases monotonically with increasing $M_{AD}$ which
coincides with   ADM mass, since there is no physical scale
except for $M_{AD}$.  This behavior means that larger $M_{AD}$
produces larger gravity  and therefore an apparent horizon is
formed even if the separation $a$  is large. On the other hand,
as shown in Fig.2,  in the case of the asymptotically de Sitter
space-time, the critical separation $a_{c}$
is not proportional to $M_{AD}$. There is a maximum value
near
$M_{AD} \simeq 0.7M_{c}$ and then the critical separation
$a_{c}$  decreases with increasing $M_{AD}$.  Further, we find
that  when an apparent horizon forms the inequality
Eq.(\ref{crit:negative})  is {\it not} satisfied, that is,
$M_{AD}$ is strictly positive.  Therefore, we may conclude as
far as we have examined that if $M_{AD} <0$ no apparent horizon
encompassing both black holes appears.

In Fig.3, $a_{c}/M_{AD}$ as a function of $H/H_{c}$ is shown,
where
$H_{c} \equiv 1/\sqrt{27}M$.
We find that $a_{c}/M_{AD}$ drops sharply near $H/H_{c} \simeq
1$.  Further, there does not appear the apparent horizon
enclosing  both black holes for all $a$ when $M_{AD}>M_{c}$ (or
$H_{AD}>H_{c}$)
 within our numerical
accuracy. This behavior essentially coincides with the
two-Einstein-Rosen bridge system investigated in \cite{nym}.
When  there is a positive cosmological constant,
it is impossible for two black holes with large $M_{AD}$ to
coalesce  and to form a single black hole.

\section{The AD Mass and Linear Momentum as a Source of
Gravity}

In the previous section, we have adopted the AD mass as a
measure  of inhomogeneities and shown its relation to
the apparent horizon formation.
However, the AD mass can vanish and could be negative
even if the inhomogeneities due to ordinary matter fields
exist.  Then, we
cannot conclude, only from the condition that AD mass vanishes,
 that such a space-time is de Sitter.

 In order to show that
$M_{AD}$ can be negative,
let us focus on the conformally flat space. Then,
we will first rewrite $\Delta M$ as
%
\beq
\Delta M
   ={H\over8\pi}\int\Bigl[{\tilde \xi}^{i}{\tilde D}_{j}
{\tilde K}_{i}^{j}
   +{1\over2}{\tilde K}^{ij}({\tilde L}{\tilde \xi})_{ij}
\Bigr]d\tilde{v}
   =H\int {\tilde \xi}^{i}{\tilde J}_{i}d\tilde{v},
\eeq
%
where the first equality comes from the Gauss's theorem and
in the  second one the use has been made of the momentum
constraint Eq.(\ref{const:mom}) and the conformal Killing
equation
$({\tilde L}{\tilde \xi})_{ij}=0$. From Eq.(3.1),
we can see that when $\xi^{i}J_{i}$ is negative,
$\Delta M$ is also negative.
Hence, in contrast with the ADM mass in the asymptotically
flat  space-time, $M_{AD}$ is reduced by the imploding motion
of matter fields.  Furthermore, in this case, $M_{AD}$ can be
negative  since it is possible to consider arbitrary large $H$
or, in other words,  the support of the integrand in Eq.(3.1)
can be arbitrary large.

Here we shall estimate the condition to make $M_{AD}$ negative.
Assuming the dominant energy condition,
$ {\rho}_{H} \geq (J_iJ^i)^{1/2}$,
we obtain
%
\beq
|\Delta M| \simeq HL \times |{\tilde J}_{r}|V
   \siml HL \times {\tilde \rho}_{H}V,
\eeq
%
where $L$ and $V$ are, respectively, the length scale and
volume scale  of the region in which inhomogeneities exist.
On the other hand, from Eqs.(\ref{laplace}) and (\ref{mass}),
we obtain
%
\beq
 M=\int\Bigl({\tilde \rho}_{H}\psi^{-3}
    +{1\over16\pi}{\tilde K}_{i}^{j}{\tilde K}_{j}^{i}\psi^{-7}
\Bigr)d\tilde{v}
    \simg {\tilde \rho}_{H}V.
\eeq
%
 From the above equation, we find the condition of $M<|\Delta
M|$ to be
%
\beq
 L \simg H^{-1}.
\eeq
%
Hence, in order that $M_{AD}$ is negative, the size of the
inhomogeneities  should be larger than the cosmological horizon
scale.

In the case of the asymptotically flat space-time,
the ADM mass is the conserved energy. Furthermore
it can be a source of gravity (tidal force) and therefore
it is called  the $gravitational$ $mass$.
As mentioned,
the AD mass is also the conserved quantity
and has a meaning of energy in the asymptotically de Sitter
space-time. However,  a question arises here:
``Does the AD mass produce the gravitational field
in the same manner as the ADM mass in
the asymptotically flat space-time?"
In order to  answer this question, by analyzing
the conformally flat initial data, we shall investigate
the asymptotic behavior of the electric part of the Weyl
tensor  which corresponds to the tidal force for timelike
geodesic
normal to the spacelike hypersurface considered here.
Here we shall
consider three cases, i.e., initial data in a spherical
symmetric space (A),  with a spherical source with a linear
momentum (B), and  with two boosted masses
of which the total momentum vanishes (C).
To obtain the $\lq\lq$physical" components, we introduce
an orthonormal triad of basis vectors $e^{i}_{(\alpha)}$ as
%
\beqa
 e^{i}_{(r)}&=&{1 \over \psi^{2}r}(x,y,z), \\
             e^{i}_{(\theta)}&=&{1  \over
\psi^{2}r\sqrt{x^{2}+y^{2}}}
                               (xz,yz,-x^{2}-y^{2}),  \\
             e^{i}_{(\varphi)}&=&{1 \over
\psi^{2}\sqrt{x^{2}+y^{2}}}
                               (-y,x,0).
\eeqa
Using Eq.(\ref{cmc}), the triad components of the electric
part of  the Weyl tensor in the vacuum region is written as
%
\beq
 E_{(\alpha)(\beta)}\equiv
C_{(\alpha)\mu(\beta)\nu}t^{\mu}t^{\nu}
    ={^{(3)}}R_{(\alpha)(\beta)}-3H{\hat K}_{(\alpha)(\beta)}
    -{\hat K}_{(\alpha)}^{i}{\hat K}_{(\beta)i},
\eeq
%
where $t^{\mu}$ is the unit vector normal to the
hypersurface.

Since we consider the conformally flat initial data,  the
asymptotic  behavior of the metric is given by
Eq.(\ref{asympt}).  Hence the triad components
of the Ricci tensor $^{(3)}R_{(\alpha)(\beta)}$ of  the 3-space
asymptotically behave as
%
\beqa
   {^{(3)}}R_{(r)(r)}& \longrightarrow &-{2M \over r^{3}},  \\
   {^{(3)}}R_{(\theta)(\theta)} &\longrightarrow&  {M \over
r^{3}}, \\
   {^{(3)}}R_{(\varphi)(\varphi)} &\longrightarrow&  {M \over
r^{3}},
\eeqa
%
and the other components are of order $O(r^{-4})$.
Thus the traceless part of
the extrinsic curvature ${\hat K}_{ij}$ gives rise to
the difference among the various conformally flat  initial
data.

\subsection{Spherically Symmetric Initial Data}

In a  spherically symmetric space, Eq.(\ref{conf:mom})
is easily integrated as
%
\beq
 {\tilde K}_{(r)(r)}=-2{\tilde K}_{(\theta)(\theta)}
   =-2{\tilde K}_{(\varphi)(\varphi)}
   \longrightarrow {S_{J} \over r^{3}},~~~{\rm for}~~r
\rightarrow +\infty
\eeq
%
where
%
\beq
 S_{J} \equiv 8\pi\int_{0}^{+\infty}r'^{3}{\tilde J}_{r}dr',
\eeq
%
and other components of ${\tilde K}_{(\alpha)(\beta)}$ vanish.
Then, it is easily seen
that for $r \rightarrow +\infty$, 
$E_{(\alpha)(\beta)}$ behaves  as
%
\beqa
E_{(r)(r)}&=&-2E_{(\theta)(\theta)}=-2E_{(\varphi)(\varphi)}
   \rightarrow -{2M \over r^{3}}-H{\tilde K}_{(r)(r)}
\nonumber\\
  &=&-{2 \over r^{3}}
    \Bigl(M+{H\over 8\pi}\oint {\tilde K}_{i}^{j}{\tilde
\xi}^{i}
    d{\tilde S}_{j} \Bigr)
   =-{2M_{AD} \over r^{3}},
\label{err}
\eeqa
%
and other components are $O(r^{-4})$.
Eq.(\ref{err}) shows that $M_{AD}$ produces the tidal force
by the same way as ADM mass in the asymptotically flat
space-time.

\subsection{A Spherical Source with Linear Momentum}

The solution of the momentum constraint (\ref{conf:mom}) for
a single black hole located at the origin is  obtained  by
keeping only
$K^{(+) ij}$ nonzero with $a=0$ in Eq. (\ref{Kij}).  Then
$P$ is a conserved linear momentum associated with the
translational  Killing vector of the background de Sitter
space-time.  Then the asymptotic behavior of the extrinsic
curvature is given by
%
\beqa
   {\tilde K}_{(r)(r)}
   &\longrightarrow& {3 \over r^{2}}P\cos\theta,   \\
   {\tilde K}_{(\theta)(\theta)} ={\tilde
K}_{(\varphi)(\varphi)}
   &\longrightarrow &-{3 \over 2r^{2}}P\cos\theta,
\eeqa
%
and the other components vanish.
We can easily verify that in this case $M_{AD}$ coincides
with $M$  considering Eqs.(\ref{admass})$\sim$(\ref{deltam}).
However, the asymptotic behavior of
$E_{(\alpha)(\beta)}$ is different from the asymptotically
flat case:
%
\beqa
   E_{(r)(r)} &\longrightarrow& -{2M_{AD} \over r^{3}}-{3H
\over r^{2}}
    P\cos\theta,  \\
   E_{(\theta)(\theta)}=E_{(\varphi)(\varphi)}
    &\longrightarrow& -{M_{AD} \over r^{3}}-{3H \over 2r^{2}}
    P\cos\theta,
\eeqa
and other components vanish.
$M_{AD}$ produces the tidal force by the same manner
as the asymptotically flat space-time.
However, in the limit of $r \rightarrow \infty$, the leading
term of
$E_{(\alpha)(\beta)}$ comes from the linear momentum $P$ and
depends on the polar angle $\theta$.\footnote{
This effect can be regarded as a kinematical effect due to 
taking the comoving coordinate system. In fact, the linear momentum
can be eliminated by a coordinate transformation. We note, however,
that the dipole term in the tidal force for a test particle cannot be
eliminated by the coordinate transformation\cite{nakao}.
}
The relative strength between the tidal force due to the
momentum $P$  and that due to $M_{AD}$ is given by $Hr \times
|P|/M_{AD}$.  Hence, in order that the tidal force due to the
linear momentum dominates, the cosmological constant should
satisfy
$ |P|/M_{AD} \simg 1$
if the inhomogeneity
distributes within a region smaller than the cosmological
horizon scale.  On the other hand, if the inhomogeneity
distributes over the cosmological  horizon scale and there is a
coherent imploding motion in addition to  the translational
motion, the $M_{AD}$ can vanish. In such a case,  the effect of
the linear momentum on the tidal force also dominates.  It
should be noted that the tidal force  is produced by the linear
momentum of a matter field even if $M_{AD}$ vanishes.

\subsection{Initial Data for Two Black Holes with Linear
Momenta}

In contrast with the previous case, the total linear  momentum
vanishes  in the case of the initial data obtained in Sec.II.
From Eqs.(\ref{def:lw}) and (\ref{wi}), the asymptotic behavior
of
${\tilde K}_{(\alpha)(\beta)}$ is
%
\beqa
   {\tilde K}_{(r)(r)}
   &\longrightarrow& {6 \over r^{3}}aP(\cos{2\theta}+1),    \\
   {\tilde K}_{(\theta)(\theta)}
   &\longrightarrow& -{3 \over 2r^{3}}aP(\cos{2\theta}+3),   \\
   {\tilde K}_{(\varphi)(\varphi)}
   &\longrightarrow& -{3 \over 2r^{3}}aP(3\cos{2\theta}+1),
\eeqa
%
and the other components are $O(r^{-4})$.
Hence $M_{AD}$ is given by
%
\beq
M_{AD}=M+2HaP,
\eeq
%
and for $r \rightarrow +\infty$, $E_{(\alpha)(\beta)}$ is given as
%
\beqa
   E_{(r)(r)}&\longrightarrow& -{2M_{AD} \over r^{3}}
    -{6 \over r^{3}}HaP(3\cos{2\theta}+1), \label{twoerr} \\
   E_{(\theta)(\theta)}&\longrightarrow& {M_{AD} \over r^{3}}
    +{1 \over 2r^{3}}HaP(3\cos{2\theta}+5),  \label{twoett}\\
   E_{(\varphi)(\varphi)}&\longrightarrow& {M_{AD} \over r^{3}}
    +{1 \over 2r^{3}}HaP(9\cos{2\theta}-1), \label{twoepp}
\eeqa
and the other components are of order $O(r^{-4})$.
Eqs.(\ref{twoerr})-(\ref{twoepp}) means that when the motion
of the inhomogeneities
is not isotropic the tidal force depends on their directions
even if the total momentum vanishes.
It is worthy to notice that even if $M_{AD}$ vanishes, the
tidal force  is indeed produced by the anisotropic motion of a
matter field   as in Sec.III.B.

\section{Discussion}

We have numerically obtained initial data of two boosted
black holes in the asymptotically de Sitter space-time. We
have   searched for an apparent horizon enclosing both two
black holes and  have shown the relation between the apparent
horizon formation and AD mass $M_{AD}$.
Interestingly, when $M_{AD}$ is larger than the critical  mass
$M_{c}$ of  the Schwarzschild-de Sitter space-time, the
apparent horizon enclosing  both black holes does not appear
for any separation $a$ between  each black hole and for any
momentum.  This result is the same as that in the analysis of
the  Einstein-Rosen bridge system \cite{nym}.

Furthermore, in order to understand the dynamical meaning  of
$M_{AD}$,  we have examined  the asymptotic behavior of the
electric part of the Weyl tensor $E_{(\alpha)(\beta)}$,
which corresponds to the tidal force, in three cases of
initial data in the asymptotically de Sitter space-time.
In the spherically symmetric case, $M_{AD}$ produces the
tidal force  by the same way as the asymptotically flat
space-time,  i.e., $M_{AD}/r^{3}$. However, in the case of the
initial data of  a spherical source with linear momentum, the
behavior of the  tidal force is very different from the
asymptotically flat case.  In this case, although $M_{AD}$
produces a part of  the tidal force by the same way as
the asymptotically flat space-time,
the leading term of $E_{(\alpha)(\beta)}$ is proportional
to $r^{-2}$, which comes from the linear momentum,
and depends on the direction. In the case of the initial  data
obtained in Sec.II, i.e.,
two boosted black holes, $M_{AD}$ also
produces
 the tidal force in the same way as the above examples but
the contribution of the relative velocity produces the
anisotropic  dependence of the tidal force on the polar angle
$\theta$.  However, in contrast with a
spherical source with linear momentum, both contributions
from
$M_{AD}$ and linear momenta on the tidal force have the same
asymptotic behavior $r^{-3}$ since the total momentum vanishes.

 From these results, it seems that the anisotropic velocity
is important for the dynamics of inhomogeneities in the scale
comparable to the cosmological horizon scale.
On the other hand, in the analysis of initial data,
we have seen that the apparent horizon formation crucially
depends on
 whether the AD mass is larger than $M_{c}$ or not.
This fact does not depend on the anisotropic velocity and
suggests that the final state of non-rotating gravitational
collapse in the asymptotically de Sitter space-time
is the Schwarzschild-de Sitter space-time if the cosmic
censorship holds. Inhomogeneities with too large $M_{AD}$
cannot collapse into a black hole again if the cosmic
censorship holds.

If there exists  a positive cosmological constant,
large scale nonspherical gravitational
collapse may be very different from that of the
asymptotically flat space-time due to the different behavior
of
$E_{(\alpha)(\beta)}$.
In order to investigate such an effect on the dynamics of the
inhomogeneities, we need to solve the Einstein equation
numerically and  follow the time evolution.

\acknowledgments
We thank I. Moss for useful discussion in an earlier stage of
this work.  We are also grateful
to H. Sato, H. Shinkai and T. Nakamura for their helpful discussion.
This work
was supported partially by a Grant-in-Aid for
Scientific Research Fund of the Ministry of
Education, Science and Culture (Specially Promoted
Research No. 08102010), by a JSPS Grant-in-Aid (No.
053769), and by the Waseda University Grant  for
Special Research Projects.

\newpage

\section*{Figure Captions}

\vspace*{12pt}
\noindent
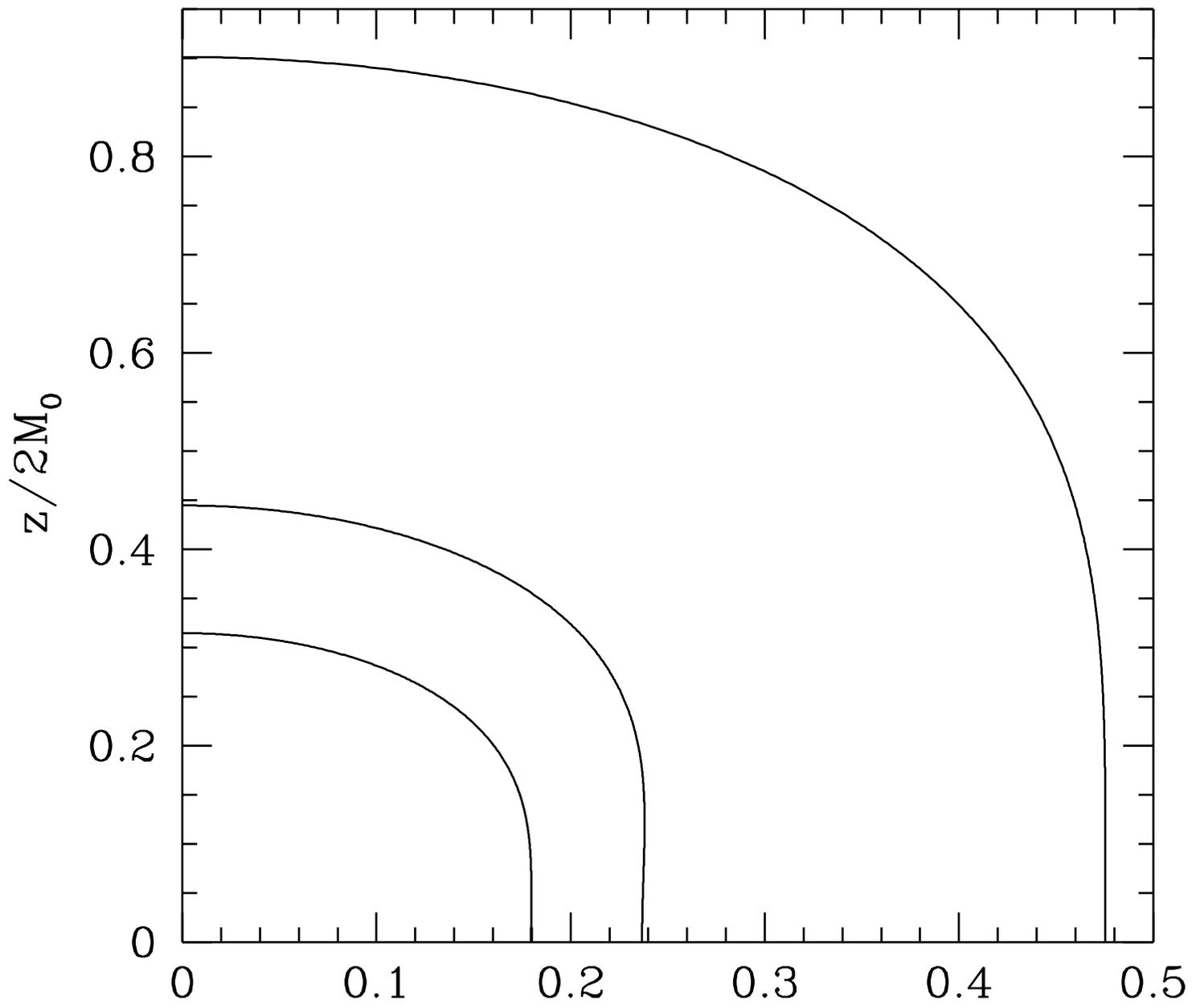
{\bf Figure 1:}
The shapes of apparent horizons near critical separations for
$P/M_0=-5$ and $H/2M_0=0,0.1,0.15$ from outside to inside.

\vspace*{12pt}
\noindent
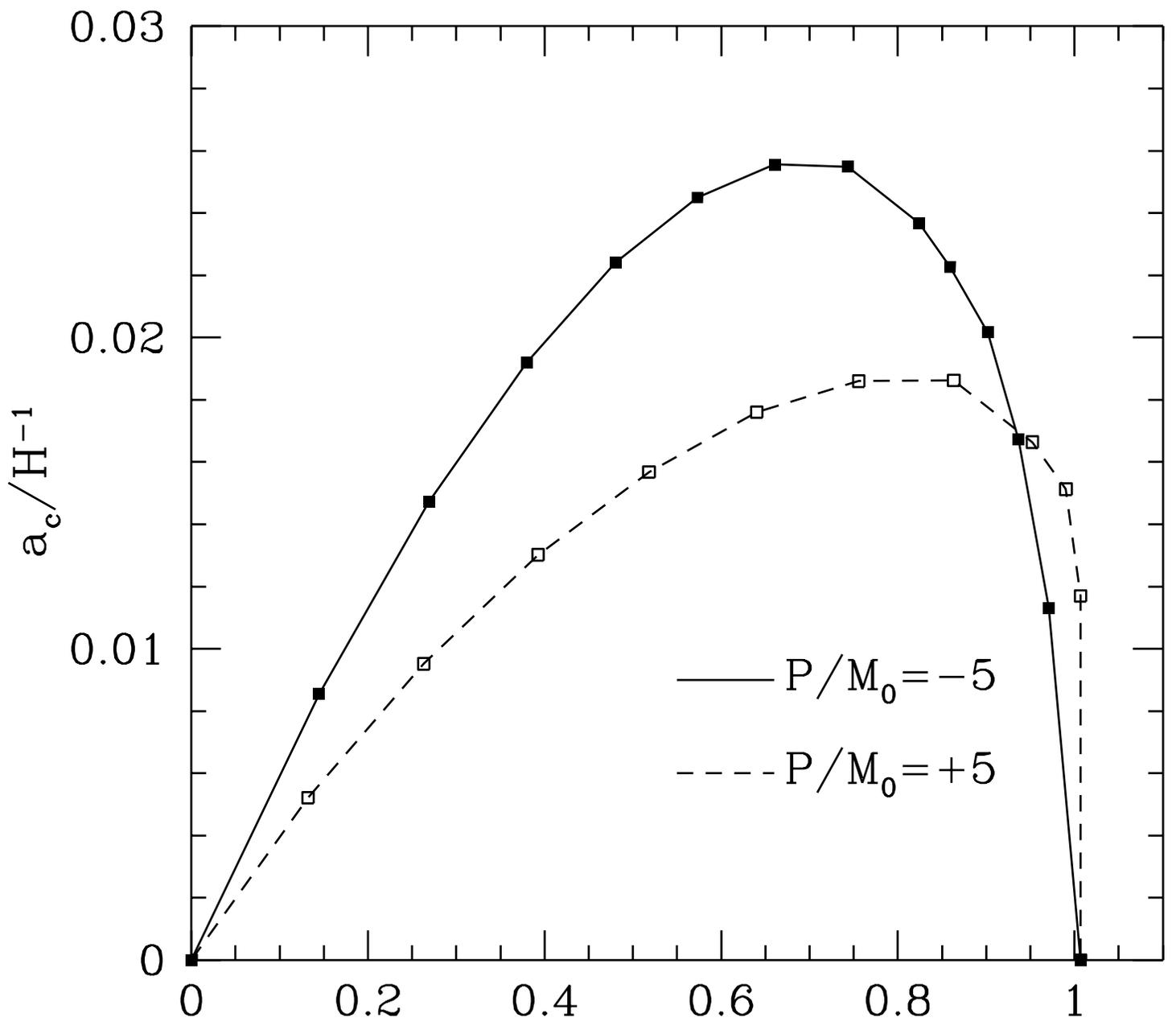
{\bf Figure 2:}
The relation between the critical separation $a_{c}/H^{-1}$
and AD mass
$M_{AD}$. (a) is the case with $|P|/M_{0}=1$. The white square
is for  the case with $P>0$ and the black square corresponds to
that of $P<0$.   (b) is the same as (a) but
with $|P|/M_{0}=5$.   The
critical value looks slightly shifted upward (by $0.7\%$)
due to a numerical error.

\vspace*{12pt}
\noindent
{\bf Figure 3:}
The relation between the critical separation $a_{c}/M_{AD}$
and cosmological constant $H/H_c=\sqrt{\Lambda /\Lambda_{c}}$.

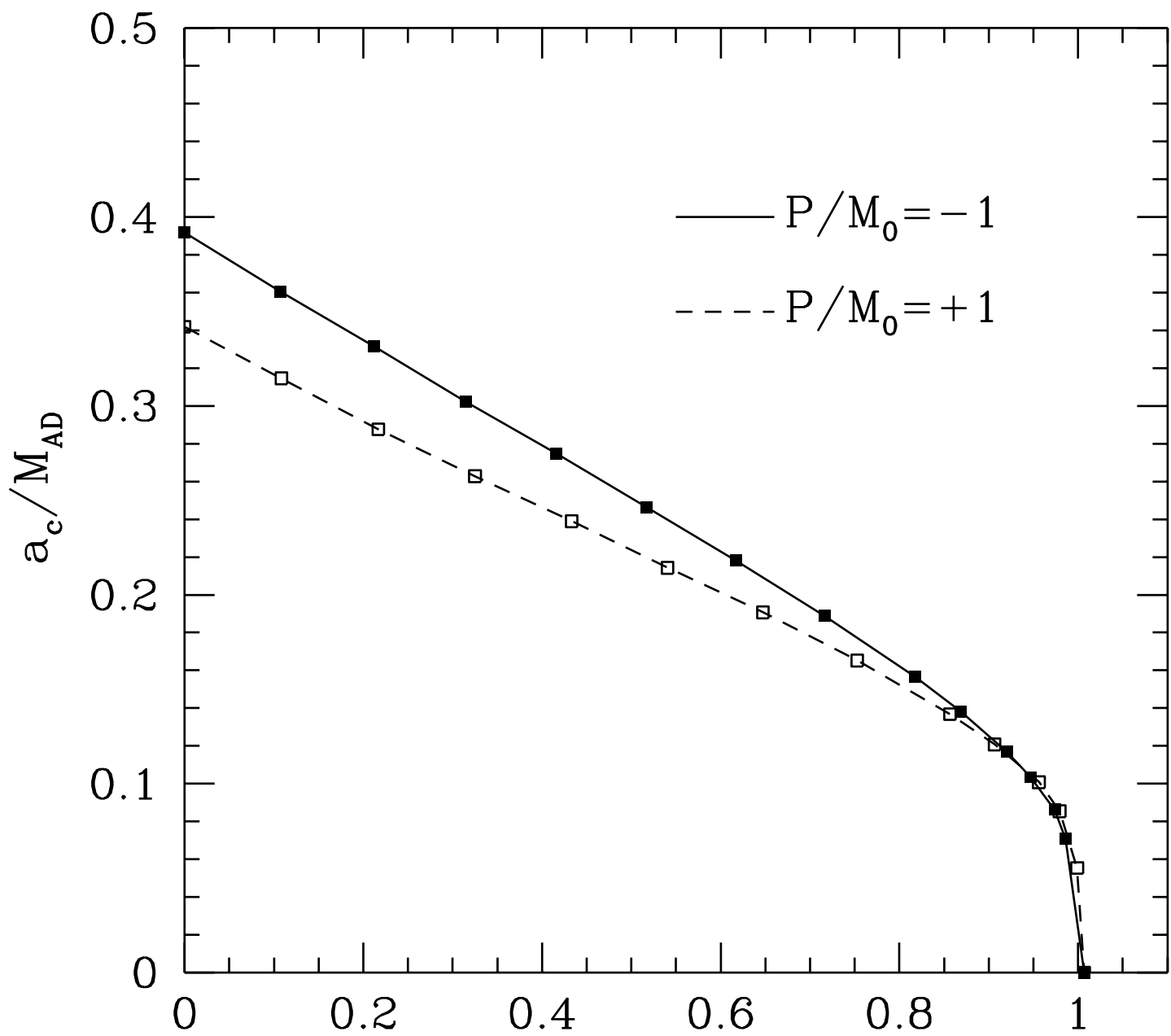
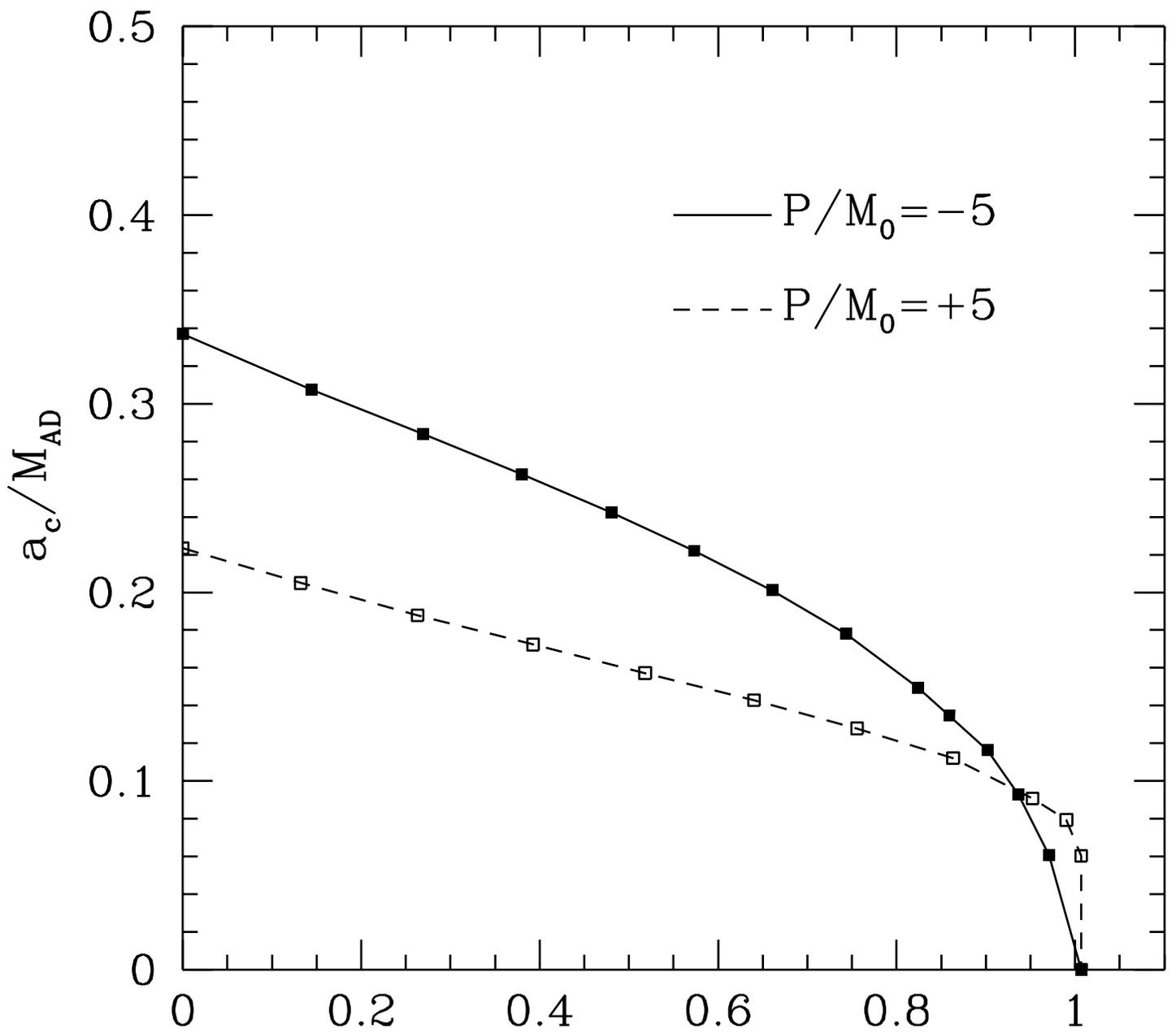
\end{document}